\documentclass[%
aps,
prl,
twocolumn,
superscriptaddress,
amsmath,
amssymb,
citeautoscript
]{revtex4-1}
\usepackage{graphicx}
\usepackage{dcolumn}
\usepackage{bm}
\usepackage{hyperref}

\usepackage{xcolor}
\usepackage{soul}

\newcommand{\RNum}[1]{\uppercase\expandafter{\romannumeral #1\relax}}

\begin{document}

\preprint{APS/123-QED}

\title{Basolateral mechanics prevents rigidity transition in epithelial monolayers}
\author{Jan Rozman}
\email{jan.rozman@physics.ox.ac.uk}
\affiliation{ Rudolf Peierls Centre for Theoretical Physics, University of Oxford, Oxford OX1 3PU, United Kingdom}
\author{Matej Krajnc}
\affiliation{ Jo\v zef Stefan Institute, Jamova 39, SI-1000 Ljubljana, Slovenia}
\author{Primo\v z Ziherl}
\affiliation{ Jo\v zef Stefan Institute, Jamova 39, SI-1000 Ljubljana, Slovenia}
\affiliation{Faculty of Mathematics and Physics, University of Ljubljana, Jadranska 19, SI-1000 Ljubljana, Slovenia}

\date{\today}
            
\begin{abstract}
The mechanics of epithelial tissues, which is governed by forces generated in various cell domains, is often investigated using two-dimensional models that account for the apically-positioned actomyosin structures but neglect basolateral mechanics. We employ a more detailed three-dimensional model to study how lateral surface tensions affect the structure and rigidity of such tissues. We find that cells are apicobasally asymmetric, with one side appearing more ordered than the other depending on cell target perimeter. In contrast to the 2D model, which predicts a rigidity transition at large target perimeters, tissues in the 3D model remain solid-like across all parameter space.
\end{abstract}

\maketitle

{\textit{Introduction.}}---Large-scale structural and material pro\-perties of tissues as well as their flow- and deformation patterns strongly depend on the mechanics at the level of cell cortex and transmembrane proteins, which mediate active contractile forces and cell-cell adhesion forces, respectively. This mechanics governs the ability of cells to flow and rearrange, which is crucial in various biological contexts including morphogenesis, wound healing, and cancer metastasis.

The mechanical behavior of epithelial tissues is most often explored using the area- and perimeter-elasti\-city~(APE) vertex model~\cite{honda1980much,Farhadifar07,fletcher2014vertex,alt2017vertex}. This model describes tissues in two dimensions (2D) by considering only the mechanics of apical cell surfaces, assuming ener\-gy penalties for deviations of perimeters and surface areas from their respective target values. Within the APE model, the intrinsic epithelial rigidity can be tuned by the target cell-shape index  $q_0$ (i.e., the dimensionless target perimeter), which measures the relative weight of contractile and adhesive forces. In particular, 
\begin{equation}\label{eq:shape_index}
    q_0=\frac{\gamma}{4\xi\sqrt{A_0}}\>,
\end{equation}
where $\gamma$ is the cell-cell adhesion strength,  $\xi$ is the active contractility of the apical cell perimeter, and $A_0$ is the target area. In the APE model, $q_0$ serves as a control parameter for the rigidity transition~\cite{Bi15}, whereby tissues switch between the rigid and the soft mode. This transition was originally reported at $q_0\approx 3.81$, which corresponds to the shape index of the regular pentagon, with further studies placing it between 3.8 and 3.9~\cite{merkel2019minimal,wang2020anisotropy,tong2022linear}. The value of $q_0$ also controls the response in the nonlinear regime where the energy barriers for local cell-neighbor exchanges may vanish, permitting cell rearrangements. Moreover, in a model where the actual cell-shape index defined by $q=P/\sqrt{A}$ ($P$ and $A$ being apical perimeter and area, respectively) is fixed and identical for all cells~\cite{Hocevar09}, there exists an order-disorder transition at $q=3.81$, whereas the measured cell-shape index in a tissue fluidized by tension fluctuations exceeds $3.81$~\cite{Krajnc20}.

The molecular machinery generating contractile and adhesive forces that underpin perimeter elasticity is located close to the apical side~{\cite{ buckley2022apical,campas2023adherens}}. While this seems to justify the 2D nature of the APE model, the mechanics of the apical sides may signi\-ficantly depend on their coupling to basola\-teral regions. Furthermore, the statistical properties of in-plane cell arrangements, which are commonly characterized apically, may be misinterpreted due to the reduced-dimensionality approximation. For instance, the mean apical surface area is related to the number of cell neighbors by the empirical Lewis law~\cite{Lewis28}, and if we assume that cells are right prisms, this law implies that their volumes are very different depending on the number of neighbours. In a model that describes cells more generally as polyhedra with their apical and basal polygons connected by four-sided lateral sides, the additional degrees of freedom may permit tight packings of cells of equal volumes but variable apical and basal surface areas. 

We address the role of basolateral regions by numerically studying  a three-dimensional (3D) vertex model of epithelial monolayers~\cite{okuda2015vertex,bielmeier2016interface,misra2016shape,alt2017vertex}, which includes the full 3D surface mechanics of cells and accounts for their inhe\-rent apico-basal asymmetry associated with the apical positioning of the force-generating machinery. We observe that due to this asymmetry, cell shapes too are asymmetric: At small target perimeters, the apical side is more ordered and reminiscent of a solid-like tissue whereas the basal side is more disordered and appears fluid-like. This is reversed as the target perimeter is increased. Crucially, we find that the coupling of the apical side to the basolateral region solidifies the tissue such that the shear modulus and the energy barriers for local neighbor exchange are finite at all target perimeters, preventing the rigidity transition observed in 2D.

\begin{figure*}[tbh!]
    \includegraphics[width=16cm]{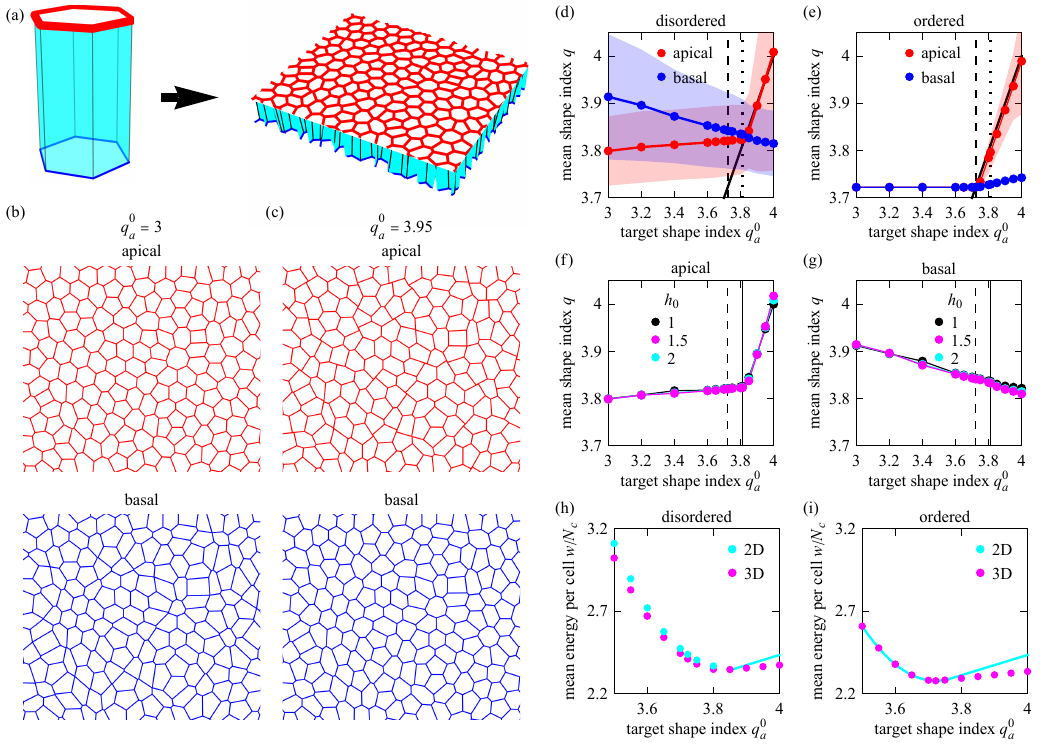}
    \caption{
    (a) Schematic of the model: The epithelial monolayer contains cells with apical perimeter elasticity, lateral surface tension, and volume elasticity. (b, c) The apical (top) and basal (bottom) sides of a part of the model tissue at $q_a^0=3$ (b) and $q_a^0=3.95$ (c). (d, e) Mean cell-shape index of the apical and basal side as a function of $q_a^0$ for a disordered~(d) and an ordered~(e) tissue. Shaded regions show standard deviation. Dashed and dotted lines indicate $q_a^0=3.72$ and $3.81$, respectively. Thick black lines show $q=q_a^0$. (f, g) Collapse of the mean apical~(f) and basal~(g) cell-shape indices for different $h_0$. (h, i)~Comparison of energies of a disordered~(h) and an ordered~(i) tissue obtained using the 2D APE model and the 3D model.}
    \label{fig:asym}
\end{figure*}

{\textit{The model.}}---We represent the tissue by a 3D packing of polyhedral cells in a monolayer, parametrizing their shapes by vertex positions $\boldsymbol r_i=(x_i,y_i,z_i)$. The apical and basal cell sides are polygons whereas the lateral sides are quadrilate\-rals~(Fig.~\ref{fig:asym}a). We describe cell mechanics by a potential energy that includes apical perimeter elasti\-city~\cite{Farhadifar07}, cell volume elasti\-city, and lateral surface ener\-gy~{\cite{derganc2009equilibrium,Krajnc18,rozman2020collective,rozman2021morphologies,okuda2020mechanical}. In dimensionless form, the energy reads
\begin{align}
    \label{eq:dimensionless}
    w = \sum_{i=1}^{N_c} \left[k_p \left(p_a^{(i)}-p_a^0\right)^2+ k_V \left(v^{(i)}-1\right)^2+\frac{1}{2}a_l^{(i)}\right],
\end{align}
where the sum runs over all $N_c$ cells. The first term represents the perimeter elasticity of the apical side associated with the actomyosin network. Here $p_a^{(i)}$ is the actual apical perimeter of $i$-th cell, $p_a^{0}$ is its target value, and $k_p$ is the corresponding modulus. The second term depends on cell volume $v^{(i)}$ and its large modu\-lus $k_V=100$ ensures that cells are nearly incompressible. The last term is the surface energy of the lateral sides of the $i$-th cell, $a_l^{(i)}$ denoting its total lateral area; the lateral surface tension is set to unity~\cite{footnote1}. 

In simulations, we affix the apical and basal vertices of cells to parallel planes separated by $h_0$, which represents a fixed cell height. As the apical and basal sides are planar, the apical and basal surface ener\-gies of the whole tissue are constant and can be omitted. As shown below, when all cells are right prisms our model is equivalent to the 2D APE model. Unless stated otherwise, $h_0=1.5$ (which corresponds to a columnar tissue), $k_p=10$, and the initial state contains $\approx 54\%$ non-hexagonal cells. As the target cell-shape index $q_0$ [Eq.~\eqref{eq:shape_index}] is a key parameter in the 2D vertex model, it is convenient to rewrite the target perimeter in Eq.~\eqref{eq:dimensionless} as $p_a^0=q_a^{0}/\sqrt{h_0}$, where $q_a^0$ is the apical target cell-shape index. The ener\-gy is minimized using an explicit Euler process; details of implementation are included in Sec.~\RNum{1} of Supplemental Material~\cite{SI}. 

\textit{Cell shapes.}---First we analyze the structure of the tissue with a disordered initial tiling. Qualitatively, we find that the shape of cells is asymmetric along the apicobasal axis: For $q^0_a\lesssim 3.81$, the apical tiling resembles a solid-like (ordered) tissue, whereas the basal tiling appears more fluid-like (disordered) [Fig.~\ref{fig:asym}(b)]. We quantify this using the measured cell-shape index $q_s^{(i)}=p_s^{(i)}/\sqrt(a_s^{(i)})$ ($s=a$ for apical and $s=b$ for basal side); here $p_s^{(i)}$ and $a_s^{(i)}$ are the perimeter and area of $i$-th cell, respectively, on the side in question [Fig.~\ref{fig:asym}(d)]. 

As target apical perimeter $p_a^0$ is increased, the diffe\-rence between the mean apical and basal cell-shape indices decreases until they become equal at a target cell-shape index $\approx3.81$. At this point, the cells are symme\-tric apicobasally (i.e., prismatic) [Fig.~{S2}(a)~\cite{SI}]. For $q_a^0\gtrsim 3.81$, the apical cell perimeter begins to follow its target value and the two measured mean indices diverge again: Now the apical shape index is the larger one and the apical side appears more fluid-like than the basal side [Fig.~\ref{fig:asym}(c),(d)]. This behavior is also seen in cuboidal and tall columnar epithelia at $h_0=1$ and 2, respectively, with the mean cell-shape indices for different $h_0$ collapsing well at a given  $q_a^0$ [Fig.~\ref{fig:asym}(f),(g)]. Figures~S{3} and S{4}~\cite{SI} show how the mean cell-shape indices depend on $k_p$ and on the extent of initial disorder in the model tissue, respectively.

The mean apical and basal shape indices behave diffe\-rently in a tissue that starts as a hexagonal lattice. For $q_a^0\lesssim 3.72$ (i.e.,~the value corresponding to the regular hexagon), the model tissue remains apicobasally symmetric, specifically a regular hexagonal lattice. Above this value of $q_a^0$ the apicobasal symmetry breaks, with the mean apical shape index again becoming larger than the basal one [Fig.~\ref{fig:asym}(e), {S2}(b),(c)~\cite{SI}].

If the cells are constrained to be apicobasally symmetric (i.e.,~if they are right prisms), the energy in Eq.~\eqref{eq:dimensionless} reduces to the usual APE energy
\begin{equation}\label{eq:APE}
    w = \sum_{i=1}^{N_c} \left[ k_p \left(p_a^{(i)}-p'_0\right)^2+ k_a \left(a_a^{(i)}-a_0\right)^2\right] +C,
\end{equation}
where $a_a^{(i)}$ is the apical (and also basal) area of $i$-th cell, $p'_0=p_a^0-h_0/(4 k_p)$, $k'_a=k_vh_0^2$, $a_0=1/h_0$, and $C=N_ck_p\left[\left(p_a^0\right)^2-\left(p'_0\right)^2\right]$. In principle, for $p'_0/\sqrt{a_0}> 3.81$ all cell areas and perimeters can assume their respective target values so that the tissue energy would equal $C$, a constant that depends on the parameters of the model but not on the geometry. Equation~\eqref{eq:APE} allows us to make a comparison between the energy minima for the equivalent 2D APE vertex model and our 3D apicobasally asymmetric tissues. For $q_a^0\gtrsim 3.81$ (3.72 for the ordered case), we can directly compare the numerically obtained energies to $C$. Below this $q_a^0$, the energy of a regular,  apicobasally symmetric hexagonal tissue can be calculated exactly. To make a comparison for the disordered tissue, we simulate a 2D APE vertex model using parameters that correspond to our 3D model as listed above. We find that in disordered tissues, the solutions of the 3D model are}energetically favorable for both small and large $q_a^0$ [Fig.~\ref{fig:asym}(h)]. The 3D ordered tissue remains symmetric and has the same energy as the 2D model for $q_a^0\lesssim  3.72$. Above this value, its energy is lower than $C$ [Fig.~\ref{fig:asym}(i)]. This shows that the apicobasally asymmetric shapes are energetically favorable compared to solutions of the 2D vertex model.

\begin{figure}[h]
    \includegraphics{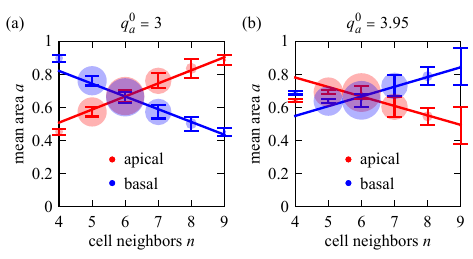}
    \caption{
    Mean apical and basal areas of model cells vs.~the number of cell neighbors, for $q_a^0=3$ (a) and $q_a^0=3.95$ (b). Lines show the Lewis law fit (excluding 4-neighbor cells), error bars indicate standard deviation, and disk areas are proportional to the frequency of cells with a given $n$.}
    \label{fig:lewis}
\end{figure}

\textit{The Lewis law.}---The empirical Lewis law~\cite{Lewis28} states that the mean area of a cell is a linear function of the number of its neighbors; since its discovery, this law was reported in many tilings seen in biological~\cite{Gomez21} as well as non-biological~\cite{Cerisier96,Prozorov08,Roth13} structures, and also in theoretical models~\cite{Farhadifar07,kokic2019minimisation}. While our model features almost incompressible cells, they are apicobasaly asymmetric due to the apical perimeter elasticity term. This allows them to agree with the Lewis law while having essentially identical volumes: For $q_a^0\lesssim 3.81$, the mean apical area of cells indeed increases linearly with the number of neighbors [Fig.~\ref{fig:lewis}(a)]. However, on the basal side a larger number of neighbors implies a smaller mean area. This can be explained as follows: A polygon with fewer sides has a larger perimeter at the same apical area. As minimizing the perimeter is energetically favorable for $q_a^0\lesssim 3.81$, cells with fewer neighbors need to have a smaller area to have the same perimeter. As a consequence, their basal areas need to increase to accommodate the same volume.  

This effect is less pronounced at large $q_a^0$ and the mean areas become almost independent of the number of neighbors at $q_a^0\approx3.81$~[Fig.~S5(a)~\cite{SI}]. At even larger target perimeters, the opposite relation comes into effect: Cells with more neighbors have smaller apical areas, whereas their basal areas are larger [Fig.~\ref{fig:lewis}(b)]. When we fit the Lewis law formula $A(n)/A_0=(n-n_0)/(6-n_0)$ ($A_0$ being the mean cell area) to our data, we find the usual values of $n_0=2$ or $3$ for the apical side at $q_a^0\approx2.1$ and $1.5$, respectively, for our default parameters [Fig.~S5(b)]. The relation between area and the number of cell neighbors does not appear to hold for 4-sided cells for $q_a^0\gtrsim 3.81$, and data for them is not included in the fits. 

\begin{figure}[t]
    \includegraphics{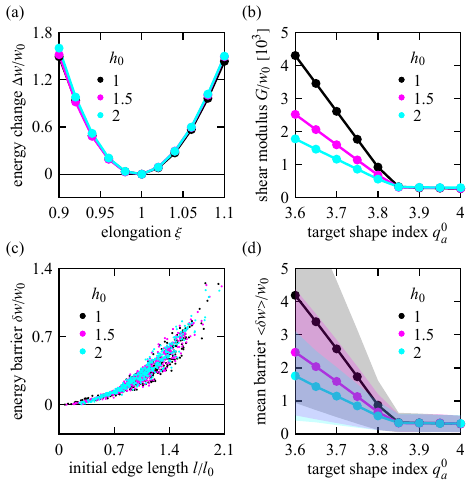}
    \caption{
    (a) Energy change upon pure shear at $q_a^0 = 3.95$. (b)~Shear modulus vs.~target shape index for three cell heights.  (c) Energy barrier for a T1 transition plotted against initial junction length for $q_a^0=3.95$: Energy barriers are always finite. (d) Mean energy barrier for a T1 transition as a function of $q_a^0$ for a disordered tissue combining data from $\sim10\%$ of junctions; shaded areas show standard deviation. Energies and the modulus in all panels are rescaled by $w_0$, whereas junction lengths are rescaled by $l_0$ as described in the text.
    }
    \label{fig:T1}
\end{figure}

\textit{Tissue rigidity.}---A hallmark feature of the 2D APE vertex model is the rigidity transition, described in the {\sl Introduction}~\cite{Bi15}. To study whether this transition still takes place when the model is extended to 3D, we measure the response of the tissue to a pure shear (stretching it along the $x$ axis by $\xi$ and compressing it along the $y$ axis by $1/\xi$; Supplemental Material Sec.~\RNum{2}~\cite{SI}). At $q_a^0=3.95$ there exists an optimal $\xi$ corresponding to an energy minimum rather than a range of deformations that would give the same minimal energy [Fig.~\ref{fig:T1}(a)], demonstrating that the tissue is intrinsically solid. To quantify this, we fit a harmonic function $\Delta w = G(\xi-\xi_0)^2/2+\widetilde{w}$, where $\Delta w$ is the change of ener\-gy after the tissue is deformed and allowed to relax at a fixed topology; as the tissue is disordered, the minimum is not precisely at $\xi=1$, requiring fitting parameters $\xi_0$ and $\widetilde{w}$. The value of $G$ is almost independent of $q_a^0$ for $q_a^0\gtrsim 3.81$, but it remains finite [Fig.~\ref{fig:T1}(b)]. By examining the separate contributions to $\Delta w$ from surface tension, perimeter elasticity, and volume elasticity, we find that the perimeter elasticity contribution to rigidity virtually vanishes for $q_a^0\gtrsim 3.81$. However, the area contribution is always finite (Fig.~{S6}~\cite{SI}).

Moreover, measuring the energy change $\delta w$ associated with the formation of four-fold vertices du\-ring a T1 transition (Supplemental Material Sec.~\RNum{2}~\cite{SI}) for $q_a^0=3.95$ reveals that it is positive on all junctions and depends approximately quadratically on the initial length of the junction~$l$. Rescaling junction lengths by the length of a junction for a regular hexagonal cell with unit volume and height~$h_0$, which is $l_0=\sqrt{2/\left(3\sqrt{3}h_0\right)}$, and all energies by the energy of the said junction $(w_0=l_0 h_0)$ collapses the data for different $h_0$ reasonably well [Fig.~\ref{fig:T1}(c)]. The same was reported in the 3D surface-tension-based vertex model without apical perimeter elasticity~\cite{Krajnc18}; the final energy after completing the T1 transition and the length of the new junction also follow the relations of the pure surface tension model as shown in Fig.~{S7}~\cite{SI}. While the mean $\delta w$ becomes almost independent of $q_a^0$ above $\approx3.81$, it does not fall to~0 [Fig.~\ref{fig:T1}(d)]. Examining the separate contributions to $\delta w$ again shows that while the the perimeter term becomes negligible to T1 barriers for $q_a^0\gtrsim  3.81$, the barrier due to the surface tension term remains finite (Fig.~{S8}~\cite{SI}). Lastly, the non-constant contributions to $G$ and $\delta w$ below $q_a^0\approx 3.81$ collapse when multiplied by $h_0$, as the perimeter term in Eq.~\eqref{eq:dimensionless} scales approximately with $1/h_0$ when recast in terms of cell shape indices~(Fig.~S9~\cite{SI}).

Altogether, the combined findings for shear response and T1 energy barriers show that while the perimeter elasticity contribution to rigidity vanishes at large target perimeters, the tissue nevertheless remains solid due to the basolateral contributions. Specifically, it becomes mechanically similar to the model based solely on lateral surface tensions~\cite{Krajnc18}, which does not feature a rigidity transition.

\textit{Discussion.}---Our theory of epithelial tissues based on the classical 2D APE vertex model extended to 3D and gene\-ralized so as to account for the contribution of basolateral mechanics produces tissues that are apicobasaly highly asymmetric. This is reflected in a finite diffe\-rence of cell-shape indices at the two sides: The apical cell sides have a smaller mean shape index than the basal ones if the target apical perimeter is below the threshold and vice versa (Fig.~\ref{fig:asym}). Furthermore, the side with the smaller shape index expresses a relationship between the number of cell neighbors and the mean area at that side that is compa\-tible with the Lewis law, whereas the opposite relation holds for the side with the larger shape index (Fig.~\ref{fig:lewis}). By inspecting its behavior under an imposed shear deformation and by measuring T1 energy barriers, we find that the tissue is always solid (Fig.~\ref{fig:T1}) and does not undergo a rigidity transition as the target perimeter is increased in contrast to  the 2D APE vertex model. In this respect, our theory differs qua\-litatively from this popular but reduced-dimensionality model, thereby emphasizing the importance of apicobasal polarization and basolateral mechanics. Naturally, although our results suggest that the apically-positioned actomyosin alone cannot soften epithelial sheets to the point of vanishing sheer modulus, tissues can nonetheless become fluid-like due to various active mechanisms that introduce local energy fluctuations such as cell division and apoptosis~\cite{Ranft10}, junctional tension fluctuations~\cite{curran2017myosin,Krajnc20,Krajnc18,Mongera18,Yamamoto22}, nematic acti\-vity~\cite{mueller2019emergence,lin2023structure,rozman2023shape,rozman2023dry}, and cell motility~\cite{Bi16,Alert20}. Finally, a passive rigidity transition can occur in bulk 3D vertex or Voronoi model tissues devoid of apicobasal polarisation~\cite{merkel2018geometrically,zhang2022topologically,sarkar2023graph,lawson2024differences}.

\begin{acknowledgments}
\emph{Acknowledgments.}---We thank Subramanian Rama\-na\-than and Rastko Sknepnek for helpful discussions. J.R. acknowledges support from the UK EPSRC (Award EP/W023849/1). The authors acknowledge financial support from the Slovenian Research Agency (research project No.~J1-3009 and research core funding No.~P1-0055).
\end{acknowledgments}

\bibliography{manuscript.bib}

\end{document}


\preprint{}

\title{
\textbf{Supplemental Material: \\ Basolateral mechanics prevents rigidity transition in epithelial monolayers}}
\author{Jan Rozman}
\affiliation{ Rudolf Peierls Centre for Theoretical Physics, University of Oxford, Oxford OX1 3PU, United Kingdom}
\author{Matej Krajnc}
\affiliation{ Jo\v zef Stefan Institute, Jamova 39, SI-1000 Ljubljana, Slovenia}
\author{Primo\v z Ziherl}
\affiliation{ Jo\v zef Stefan Institute, Jamova 39, SI-1000 Ljubljana, Slovenia}
\affiliation{Faculty of Mathematics and Physics, University of Ljubljana, Jadranska 19, SI-1000 Ljubljana, Slovenia}

\date{\today}

{
\let\clearpage\relax
\maketitle
}

\section{Model details}

{\bf Energy function.} The dimensionful energy of our model tissue reads
\begin{align}
    W = \sum_{i=1}^{N_{\rm c}} \left[\frac{\Gamma_l}{2}A_l^{(i)}+K_P  \left(P_a^{(i)}-P_a^0\right)^2+K_V  \left(V^{(i)}-V_c\right)^2\right],\>
    \label{eq:energy}
\end{align} 
where the sum is over all $N_c$ cells. A harmonic energy term with bulk modulus $K_V$ ensures an effective incompressibility of cells, with $V^{(i)}$ representing the current volumes the volume of the $i$-th cell, whereas $V_c$ is the target volume. The actomyosin network is modelled as a perimeter elasticity term that is associated only with the apical side. Therefore, $P_a^{(i)}$ is the apical perimeter of the $i$-th cell, $P_a^{0}$ is the target apical perimeter, and $K_P$ is the corresponding elasticity modulus. The lateral  surface tension is $\Gamma_{l}$, with $A_l^{(i)}$ being the total  lateral area of the $i$-th cell; the factor of $1/2$ in the lateral term arises from each lateral side being shared between two cells. We assume that the individual vertices ${\rm \bf R}_j=(x_j,y_j,z_j)$ obey the overdamped equation of motion
\begin{equation}
    \eta\dot{{\rm \bf R}}_j=-\nabla_j W\>,
    \label{eq:eqMotion}
\end{equation}
where $\eta$ is an effective viscosity associated with the movement of vertices and $\nabla_j$ is the gradient with respect to the position of the $j$-th vertex, ${\rm \bf R}_j$. To calculate the surface areas and volumes of cells, their sides are discretized into triangular facets defined by two adjacent vertices and the center (mean of all vertices) of a cell side. During the simulation, we affix the apical and basal vertices of the cells to parallel flat planes, separated by a fixed tissue height~$H_0$.

To render the energy function dimensionless, we divide it by $\Gamma_l V_c^{2/3}$ so that the final dimensionless energy reads
\begin{align}
    w = \sum_{i=1}^{N_c} \left[\frac{1}{2}a_l^{(i)}+ k_p \left(p_a^{(i)}-p_a^0\right)^2+ k_V \left(v^{(i)}-1\right)^2\right],\label{eq:dimensionless}
\end{align}
where  $a_l^{(i)}=A_l^{(i)}/V_c^{2/3}$  is the reduced lateral area, $p_a^{(i)}=P_a^{(i)}/V_c^{1/3}$ is the reduced apical perimeter, $p_a^0=P_a^0/V_c^{1/3}$ is the reduced target apical perimeter, $k_p=K_p/\Gamma_l$ is the apical perimeter elasticity modulus, $v^{(i)} = V^{(i)}/V_c$, is the reduced cell volume, and $k_V = K_V V_c^{4/3}/\Gamma_l$ is the reduced bulk modulus. The reduced tissue height then reads $h_0=H_0/V_c^{1/3}$. The corresponding equation of motion is
\begin{equation}
    \dot{{\rm \bf r}}_j=-\nabla_j w\>,
    \label{eq:eqMotion}
\end{equation}
${\rm \bf r}_j$ being the dimensionless position of the $j$-th vertex measured in units of $V_0^{1/3}$. Here $\nabla_j$ indicates the gradient relative to ${\rm \bf r}_j$ and simulation time is measured in units of $\tau_0=\eta/\Gamma_l$. We use a 1024 cell sheet with periodic boundary conditions; the size and shape of the simulation box corresponds to that of a 32 by 32 cell regular honeycomb tissue with unit cell volume and height $h_0$. An explicit Euler scheme is used to solve the resulting equations of motion, with time step $\Delta t = 0.001$. Model tissues are relaxed for $t=1000$.

\begin{figure}[h]
		\centering
		\includegraphics{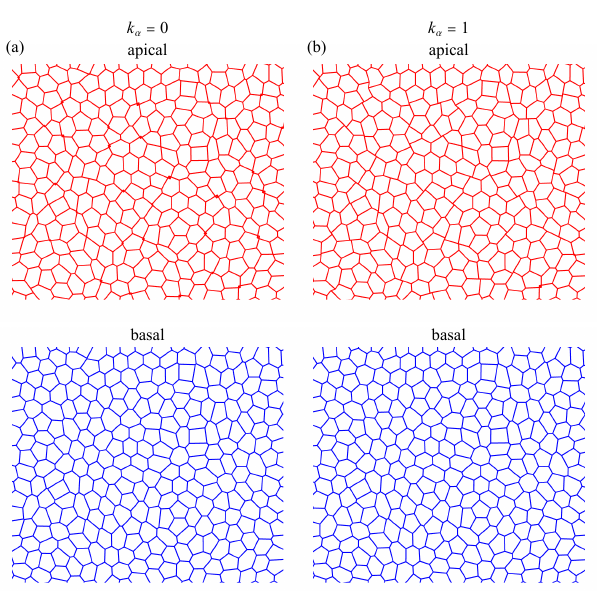}
		\caption{Example of a part of a model tissue's apical and basal sides with $q_a^0=3.95$ without the auxiliary small-angle-repulsion term [Eq.~\eqref{eq:angle}] (a) and with the term included (b), which prevents most instances of self-overlap.}
        \label{sfig:overlap}
\end{figure}

{\bf Self-overlap.} For large enough $q_a^0$, we find that the model tissue starts to self-overlap on the apical side, resulting in a non-physical configuration [Fig.~\ref{sfig:overlap}(a)]. To avoid this, we introduce an auxiliary term to the energy
\begin{equation}\label{eq:angle}
    w_\alpha=\sum_{i=1}^{N_c}\sum_{k=1}^{n^{(i)}}k_\alpha\mathrm{H}\left(\alpha_0-\alpha^{(k,k+1)}\right)\cos\left(\alpha^{(k,k+1)}\right),
\end{equation}
where the first sum is over all cells, the second is over all $n^{(i)}$ apical edges of the $i$-th cell, $\alpha^{(k,k+1)}$ is the angle between vectors along two adjacent apical edges, both starting from their shared vertex, $\mathrm{H}$ is the Heaviside function, $\alpha_0=0.05$ is a threshold angle, and $k_\alpha=1$ is the modulus. Effectively, this works as a repulsive potential preventing very small angles between apical edges, which are the precursors of apical self-overlap. This term removes most cases of self-overlap [Fig.~\ref{sfig:overlap}(b)]. The angle term contribution is not included in the measured energy in Fig.~1(h),(i). It is also not included in the dynamics when measuring shear response or T1 energy barriers.

{\bf T1 transitions.} The topology of the tissue is not fixed but can change through T1 transitions: If the length of a junction, defined as the average of the corresponding apical and basal edge lengths, falls below the threshold value 0.01 (and has decreased in length since the previous time step), it is first merged into a four-fold vertex. After a short waiting period $t_4=0.02$ it is then resolved, concluding the T1 transition: The basal vertices are separated by a distance of 0.001, whereas the new apical vertices are positioned relative to their basal partners at the same direction and distance as separated the fourfold apical and basal vertices before the resolution. A limitation of our model is that the apical and basal topology must be the same, so that scutoid cell shapes~\cite{GomezGalvez18} are not possible.

{\bf Disordered tissues.} To generate a disordered tissue with different numbers of cell neighbors starting from a regular hexagonal lattice, we temporarily add an additional fluctuating line-tension term to the energy function~\cite{curran2017myosin,Krajnc20,Rozman21}. In dimensionless form, it reads:
\begin{equation}
    w_l=\sum_{k=1}^{N_e}\gamma_k(t)\left( l_a^{(k)} + l_b^{(k)}\right)\>.
    \label{eq:activeForces}
\end{equation}
Here the sum goes over all $N_e$ cell-cell junctions, $\gamma_k(t)$ is the instantaneous apical and basal line-tension on the $k$-th junction measured in units of $\Gamma_l V_c^{1/3}$, whereas $l_a^{(k)}$ and $l_b^{(k)}$ are apical and basal edge lengths corresponding to the $k$-th cell-cell junction, measured in units of $V_c^{1/3}$, respectively. The fluctuations of the line tensions $\gamma_k(t)$ are described by the Ornstein-Uhlenbeck process
\begin{equation}
    \frac{{\rm d}\gamma_k(t)}{{\rm d}t}= -\frac{1}{\tau_m}\gamma_k(t) +\xi_k(t)\>.
\end{equation}
Here $1/\tau_m$ is the turnover rate of molecular motors (measured in units of $1/\tau_0$) and $\xi_k(t)$ is the Gaussian white noise with properties $\left <\xi_k(t)\right >=0$ and $\left <\xi_k(t)\xi_{k'}(t')\right>=(2\sigma^2/\tau_m)\delta_{k,k'}\delta(t-t')$, with $\sigma^2$ being the long-time variance of tension fluctuations (measured in units of $\Gamma_l^2 V_c^{2/3}\tau_0^{-2}$). At large enough $\sigma$, the cells can escape from their local neighborhood through a series of T1 transitions: The tissue fluidizes and becomes disordered, with a larger $\sigma$ leading to a larger fraction of non-hexagonal cells. 

The simulations run until $t=1000$ with tension fluctuations (during this period, the apical perimeter elasticity is not included in model dynamics, i.e.~$k_p=0$). The resulting model tissue is then used as our initial condition. We set $\tau_m=1$ and $\sigma=0.35$ unless specified otherwise (resulting in $\approx 54\%$ non-hexagonal cells) and generate the tissue with $h_0=1.5$; we then rescale these initial conditions for other values of $h_0$. For the ordered initial condition, we instead set $\sigma=0.05$ (too small to allow for rearrangements) so that the starting lattice is slightly perturbed.

\section{Measuring shear response and T1 energy barriers}

We conduct the following simulations on a smaller model tissue with 144 cells: We first allow the tissue to relax for $t=1000$ (for the disordered case, we start from a tissue with $\approx 49\%$ non-hexagonal cells). We then resolve all 4-fold vertices and relax the tissue for a further $t=1000$ at a fixed topology. For shear response, the tissue is then deformed as described in the main text and then relaxed for another $t=1000$, again at a fixed topology. For T1 measurements, we contract an edge into a 4-fold vertex simultaneously on the apical and basal side and record the change in energy $\delta w$ after the tissue relaxes again for  $t=1000$ at a fixed topology. The 4-fold vertices are then resolved in the opposite configuration, and the tissue relaxed for another $t=1000$ at a fixed topology to obtain data for Fig.~\ref{sfig:resolved}. 

\begin{figure}[h]
		\centering
		\includegraphics{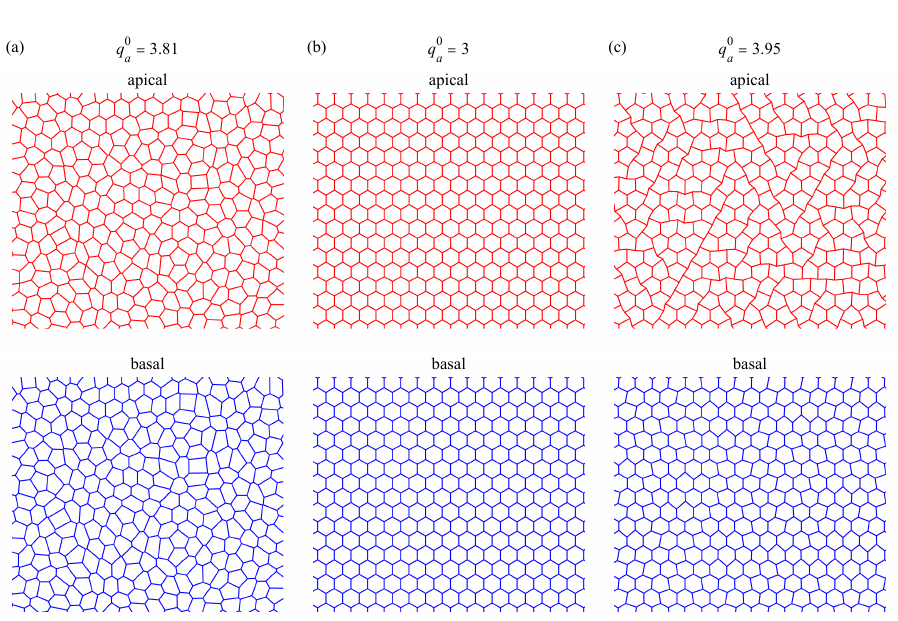}
		\caption{(a) The apical (top) and basal (bottom) sides of a part of the model tissue at $q_a^0=3.81$ for an initially disordered tissue.  (b,c) The apical (top) and basal (bottom) sides of a part of the model tissue at $q_a^0=3$ (b) and $q_a^0=3.95$ (c) for an initial ordered tissue. Tissues in panels b and c were obtained without the auxiliary small-angle-repulsion term [Eq.~\eqref{eq:angle}].}
        \label{sfig:ordered}
\end{figure}

When simulating tissue for shear response and T1 transition  measurements, we use $\Delta t=0.0001$ and we do not employ the auxiliary angle term [Eq.~\eqref{eq:angle}] to avoid including additional effects into tissue mechanics. This means that some cells may overlap as seen in Fig.~\ref{sfig:overlap}. However, since the tissue is also always solid for an ordered initial condition (Fig.~\ref{sfig:ordered_barrier}) where the overlaps are not present [Fig.~\ref{sfig:ordered}(b)], it is unlikely that they contribute significantly to the reported results. For the mean energy barriers shown in Fig.~3(d) and \ref{sfig:contributions}(g-i), the value is averaged over $\sim 10\%$ (44 out of 432) junctions in the model tissue.

\newpage
\section*{Supplementary figures}
	
\begin{figure}[h]
		\centering
		\includegraphics{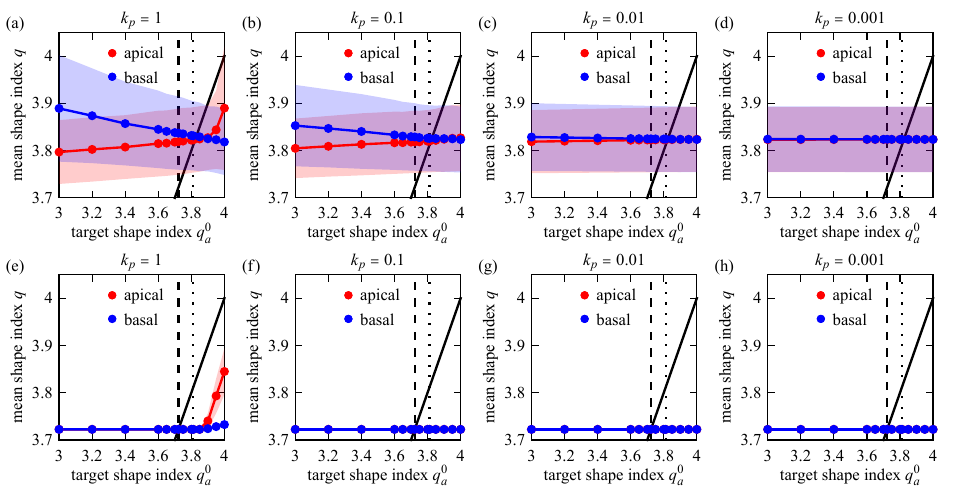}
		\caption{Mean cell-shape index of the apical and basal side of the model tissue as a function of $q_a^0$ for a disordered (a-d) and an ordered (e-f) model tissue for $k_p$ between 1 and 0.001. Values of $k_p$ are shown above the corresponding panels. Dashed and dotted lines in all panels indicate $q_a^0=3.72$ and $3.81$, respectively; thick black lines show $q=q_a^0$.}
\end{figure}

\begin{figure}[h]
		\centering
		\includegraphics{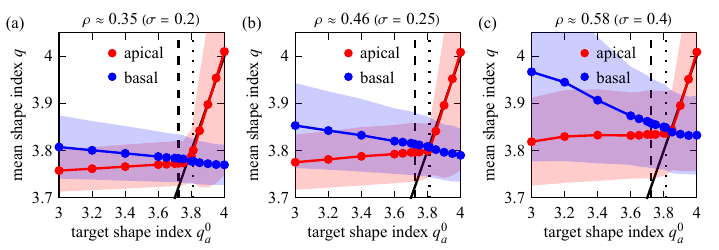}
		\caption{Mean cell-shape index of the apical and basal side of the model tissue as a function of $q_a^0$ for different levels of initial fraction of non-hexagonal cells $\rho$: $\rho\approx 0.35$ (a), $\rho\approx 0.46$ (b), and $\rho\approx 0.58$ (c). Values of $\sigma$ used to generate the initial disordered tiling are also shown above the corresponding panels in brackets. Dashed and dotted line in all panels indicate $q_a^0=3.72$ and $3.81$, respectively; thick black lines show $q=q_a^0$.}
\end{figure}

\begin{figure}[h]
		\centering
		\includegraphics{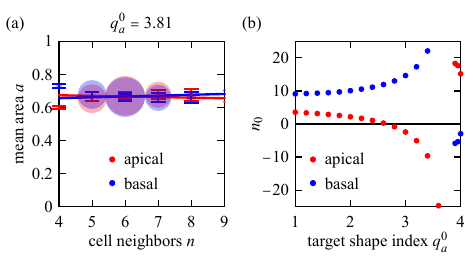}
		\caption{(a) Mean apical and basal areas of model cells as a function of the number of cell neighbors for $q_a^0=3.81$. (b) $n_0$ of the Lewis law fit as a function of $q_a^0$.}
\end{figure}

\begin{figure}[h]
		\centering
		\includegraphics{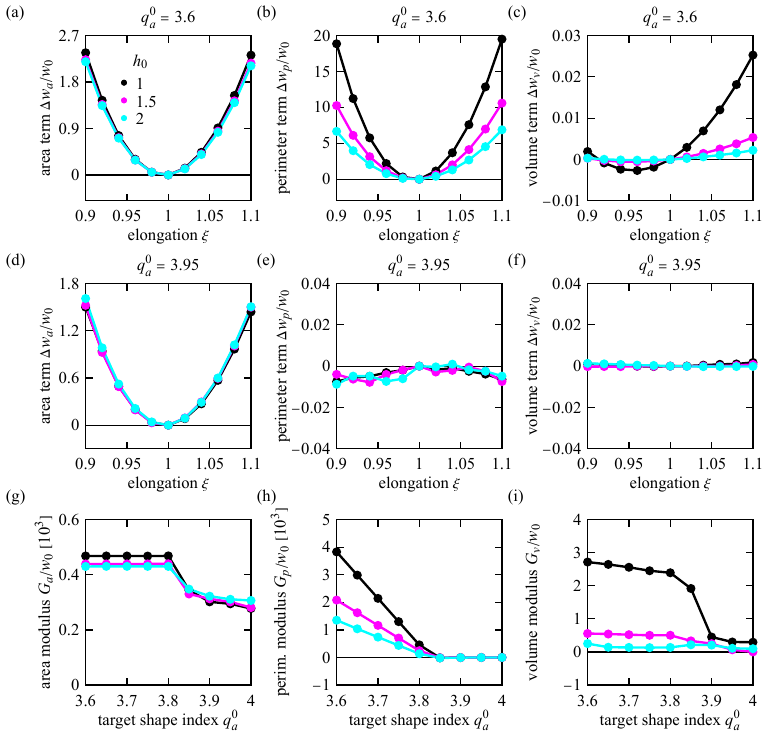}
		\caption{(a-c) Contribution to the energy change $\Delta w$ after  pure shear of the tissue  at $q_a^0 = 3.6$ from lateral area surface tensions~(a), perimeter elasticity (b), and volume elasticity (c). (d-f) Contribution to the energy change $\Delta w$ after  pure shear of the tissue $q_a^0 = 3.95$ from lateral area surface tensions (d), perimeter elasticity (e), and volume elasticity (f). (g-i)~Moduli obtained by fitting a harmonic function to the individual contributions of the energy change from pure shear arising from lateral surface tensions (g), perimeter elasticity (h), and volume elasticity (i). Energies and moduli in all panels are rescaled by $w_0$, whereas lengths are rescaled by $l_0$ as explained in the main text. Note that at large $q_a^0$, the perimeter and the volume terms are orders of magnitude smaller than the area term as shown in panels e and f, respectively. Their dependence on elongation~$\xi$ is not neatly harmonic at those $q_a^0$ as the magnitudes of these terms evidently reach the limit of the numerical accuracy. The corresponding moduli essentially vanish as shown in panels h and i (making the direct values from harmonic fits inexact); note that the scale of $G_v$ in panel i is three orders of magnitude smaller than that of $G_a$ and $G_p$ in panels g and h, respectively.}
\end{figure}

\begin{figure}[h]
		\centering
		\includegraphics{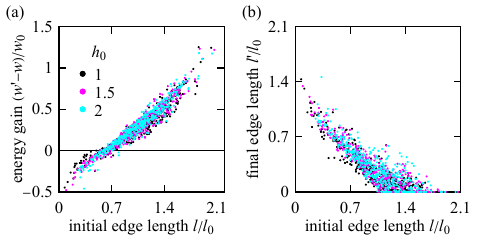}
		\caption{Rescaled final energy change $(w'-w)/w_0$ after a four-fold vertex is resolved (a) and rescaled final junction length $l'/l_0$ (b) plotted against initial junction length $l/l_0$ for $q_a^0=3.95$.}
        \label{sfig:resolved}
\end{figure}

\begin{figure}[h]
		\centering
		\includegraphics{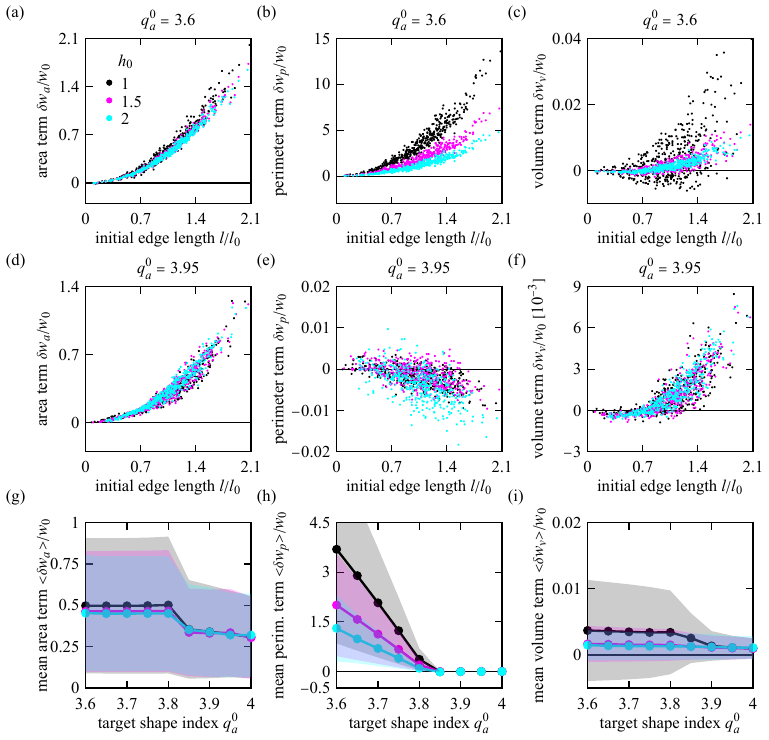}
		\caption{(a-c) Contribution to the T1 energy barrier $\delta w$ at $q_a^0 = 3.6$ from lateral area surface tensions (a), perimeter elasticity~(b), and volume elasticity (c). (d-f) Contribution to the T1 energy barrier $\delta w$ at $q_a^0 = 3.95$ from lateral area surface tensions (d), perimeter elasticity (e), and volume elasticity (f). (g-i) Mean contribution to the T1 energy barrier from lateral area surface tensions (g), perimeter elasticity (h), and volume elasticity (i), averaging over $\sim 10\%$ of junctions; shaded region show standard deviation. Energies and lengths in all panels are rescaled by $w_0$ and $l_0$, respectively, as explained in the main text.}
        \label{sfig:contributions}
\end{figure}

\begin{figure}[h]
		\centering
		\includegraphics{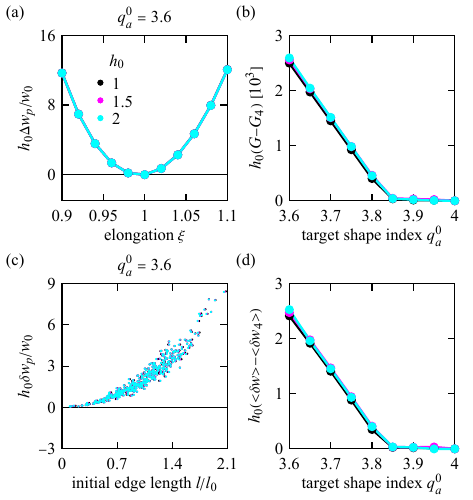}
		\caption{(a) Contribution to the energy change $\Delta w$ after  pure shear of the tissue  at $q_a^0 = 3.6$ from apical perimeter elasticity. (b) Moduli obtained by fitting a harmonic function to the energy change from pure shear with the value at $q_a^0=4$ subtracted to remove the target perimeter independent contribution. (c) Contribution to the T1 energy barrier $\delta w$ at $q_a^0 = 3.6$ from  perimeter elasticity. (d) Mean energy barrier for a T1 transition as a function of $q_a^0$  combining data from $\sim10\%$ of junctions, with the value at $q_a^0=4$ subtracted to remove the target perimeter independent contribution. Junction lengths in panel (c) are rescaled by $l_0$. Energy changes and shear moduli in all panels are multiplied by $h_0$. The resulting collapse can be explained by rewriting the perimeter term in Eq.~(2) of the main text as $k_p/h_0\left({\tilde{q}_a^{(i)}}-q_a^0\right)^2$, where $\tilde{q}_a^{(i)}=p_a^{(i)}\sqrt{h_0}$ is an estimate for the apical shape index, assuming $h_0a_a^{(i)}\approx 1$.}
\end{figure}

\begin{figure}[h]
		\centering
		\includegraphics{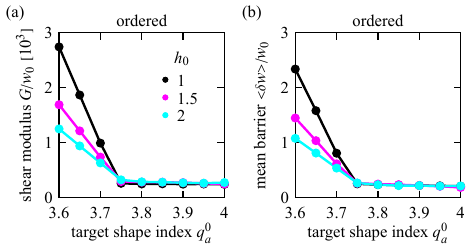}
		\caption{Rescaled modulus of the harmonic function fitted to the energy change after pure shear for an ordered initial tissue. (b) Rescaled mean energy barrier for a T1 transition as a function of $q_a^0$ for an ordered initial tissue combining data from $\sim10\%$ of junctions.}
        \label{sfig:ordered_barrier}
\end{figure}

\clearpage

\bibliographystyle{apsrev4-1}
\bibliography{supplemental.bib}